\documentclass[prc,twocolumn,superscriptaddress,showpacs,twoside,floatfix]{revtex4-2}

\usepackage{amssymb,epsfig}
\usepackage{color}
\usepackage{mathtools}
\usepackage[normalem]{ulem}

\begin{document}

\title{Mirror Symmetry Breaking Disclosed in the Decay of Three-Proton Emitter $^{20}$Al}

\author{X.-D.~Xu}
\email{Contact author: xiaodong.xu@impcas.ac.cn}
\affiliation{Institute of Modern Physics, Chinese Academy of Science, Lanzhou 730000, China}
\affiliation{School of Nuclear Science and Technology, University of Chinese Academy of Sciences, Beijing 100049, China}
\affiliation{GSI Helmholtzzentrum f\"{u}r Schwerionenforschung GmbH, 64291 Darmstadt, Germany}

\author{I.~Mukha}
\email{Contact author: I.Mukha@gsi.de}
\affiliation{GSI Helmholtzzentrum f\"{u}r Schwerionenforschung GmbH, 64291 Darmstadt, Germany}

\author{J. G.~Li}
\affiliation{Institute of Modern Physics, Chinese Academy of Science, Lanzhou 730000, China}
\affiliation{School of Nuclear Science and Technology, University of Chinese Academy of Sciences, Beijing 100049, China}

\author{S. M.~Wang}
\affiliation{Key Laboratory of Nuclear Physics and Ion-beam Application (MOE), Institute of Modern Physics, Fudan University, Shanghai 200433, China}
\affiliation{Shanghai Research Center for Theoretical Nuclear Physics, NSFC and Fudan University, Shanghai 200438, China}

\author{L.~Acosta}
\affiliation{Instituto de Estructura de la Materia, CSIC, 28006, Madrid, Spain}
\affiliation{Instituto de F\'isica, Universidad Nacional Aut\'onoma de M\'exico, A.P. 20-364, Mexico City 01000, Mexico}

\author{M.~Bajzek}
\affiliation{GSI Helmholtzzentrum f\"{u}r Schwerionenforschung GmbH, 64291 Darmstadt, Germany}
\affiliation{II.Physikalisches Institut, Justus-Liebig-Universit\"at, 35392 Gie{\ss}en, Germany}
\affiliation{Faculty of Science, University of Zagreb, 10000 Zagreb, Croatia}

\author{E.~Casarejos}
\affiliation{CINTECX, Universidade de Vigo, E-36310 Vigo, Spain}

\author{D.~Cortina-Gil}
\affiliation{Universidade de Santiago de Compostela, 15782 Santiago de Compostela, Spain}

\author{J.~M.~Espino}
\affiliation{Department of Atomic, Molecular and Nuclear Physics, University of Seville, 41012 Seville, Spain}

\author{A.~Fomichev}
\affiliation{Flerov Laboratory of Nuclear Reactions, JINR, 141980 Dubna, Russia}

\author{H.~Geissel}
\email{Deceased.}
\affiliation{GSI Helmholtzzentrum f\"{u}r Schwerionenforschung GmbH, 64291 Darmstadt, Germany}
\affiliation{II.Physikalisches Institut, Justus-Liebig-Universit\"at, 35392 Gie{\ss}en, Germany}

\author{J.~G\'{o}mez-Camacho}
\affiliation{Department of Atomic, Molecular and Nuclear Physics, University of Seville, 41012 Seville, Spain}

\author{L.V.~Grigorenko}
\affiliation{Flerov Laboratory of Nuclear Reactions, JINR, 141980 Dubna, Russia}
\affiliation{National Research Nuclear University ``MEPhI'', 115409 Moscow, Russia}
\affiliation{National Research Centre ``Kurchatov Institute'', Kurchatov square 1, 123182 Moscow, Russia}

\author{O.~Kiselev}
\affiliation{GSI Helmholtzzentrum f\"{u}r Schwerionenforschung GmbH, 64291 Darmstadt, Germany}

\author{A.A.~Korsheninnikov}
\affiliation{National Research Centre ``Kurchatov Institute'', Kurchatov square 1, 123182 Moscow, Russia}

\author{D.~Kostyleva}
\affiliation{GSI Helmholtzzentrum f\"{u}r Schwerionenforschung GmbH, 64291 Darmstadt, Germany}

\author{N.~Kurz}
\affiliation{GSI Helmholtzzentrum f\"{u}r Schwerionenforschung GmbH, 64291 Darmstadt, Germany}

\author{Yu.A.~Litvinov}
\affiliation{GSI Helmholtzzentrum f\"{u}r Schwerionenforschung GmbH, 64291 Darmstadt, Germany}

\author{I.~Martel}
\affiliation{University of Huelva, 21007 Huelva, Spain}

\author{C.~Nociforo}
\affiliation{GSI Helmholtzzentrum f\"{u}r Schwerionenforschung GmbH, 64291 Darmstadt, Germany}

\author{M.~Pf\"{u}tzner}
\affiliation{Faculty of Physics, University of Warsaw, 02-093 Warszawa, Poland}
\affiliation{GSI Helmholtzzentrum f\"{u}r Schwerionenforschung GmbH, 64291 Darmstadt, Germany}

\author{C.~Rodr\'{i}guez-Tajes}
\affiliation{Universidade de Santiago de Compostela, 15782 Santiago de Compostela, Spain}

\author{C.~Scheidenberger}
\affiliation{GSI Helmholtzzentrum f\"{u}r Schwerionenforschung GmbH, 64291 Darmstadt, Germany}
\affiliation{II.Physikalisches Institut, Justus-Liebig-Universit\"at, 35392 Gie{\ss}en, Germany}
\affiliation{Helmholtz Research Academy Hesse for FAIR (HFHF), GSI Helmholtz Center for Heavy Ion Research, Campus Gießen, 35392 Gießen, Germany}

\author{M.~Stanoiu}
\affiliation{IFIN-HH, Post Office Box MG-6, Bucharest, Romania}

\author{K.~S\"{u}mmerer}
\affiliation{GSI Helmholtzzentrum f\"{u}r Schwerionenforschung GmbH, 64291 Darmstadt, Germany}

\author{H.~Weick}
\affiliation{GSI Helmholtzzentrum f\"{u}r Schwerionenforschung GmbH, 64291 Darmstadt, Germany}

\author{P.J.~Woods}
\affiliation{University of Edinburgh, EH1 1HT Edinburgh, United Kingdom}

\author{M.V.~Zhukov}
\affiliation{Department of Physics, Chalmers University of Technology, S-41296 G\"oteborg, Sweden}

\date{\today}

\begin{abstract}
	
The previously-unknown nucleus $^{20}$Al has been observed for the first time by detecting its in-flight decays. Tracking trajectories of all decay products with silicon micro-strip detectors allowed for a conclusion that $^{20}$Al is unbound with respect to three-proton (3\textit{p}) emission. The 3\textit{p}-decay energy of $^{20}$Al ground state has been determined to be 1.93($^{+0.11}_{ -0.09}$) MeV through a detailed study of angular correlations of its decay products, $^{17}$Ne+\textit{p}+\textit{p}+\textit{p}. This value is much smaller in comparison with the predictions inferred from the isospin symmetry by using the known energy of its mirror nucleus $^{20}$N, which indicates a possible mirror-symmetry violation in the structure of 3\textit{p} emitters. 
{Such an isospin symmetry breaking is supported by the calculations of the continuum embedded theoretical frameworks, describing the observed $^{20}$Al ground state as an 1\textit{p} $s$-wave state with a spin-parity of 1$^-$, which contradicts to the spin-parity (2$^-$) of the $^{20}$N ground state.
} 
The $^{20}$Al ground state decays by sequential 1\textit{p}-2\textit{p} emission via intermediate ground state of $^{19}$Mg, which is the first observed case of ``daughter'' two-proton (2\textit{p}) radioactivity following 1\textit{p} decay of the parent state.

\end{abstract}

\maketitle

Nuclear structure beyond the proton dripline was addressed in a number of recent experimental and theoretical studies of the light- and intermediate- mass nuclei, see e.g.~the recent review in Ref.~\cite{Pfutzner:2023}. The current research status can be summarized as follows:
\\ 
i)
All known 1\textit{p} and 2\textit{p} emitters are located by 1--2 atomic mass units (\textit{amu}) beyond the proton dripline. The  2\textit{p} emitters exhibit three main decay mechanisms (direct, sequential and democratic) and their transition modes~\cite{Golubkova:2016}.
\\
ii)
The most exotic nuclei located in the very remote outskirts of the nuclear landscape become unbound in respect of new decay channels. Such exotic decay modes play an increasingly important role as the precursor's decay energy grows. The most-remote isotopes are identified as far as 4 \textit{amu} beyond the proton dripline and decay by emission of 3 or 4 protons.
\\
iii) 
The studied 3\textit{p}- and 4\textit{p}- decays show sequential decay mechanisms like 1\textit{p}--2\textit{p} and 2\textit{p}-2\textit{p} emissions, respectively. In particular, there are several isotopes whose ground states (g.s.) are established as 3\textit{p} emitters, i.e. $^{7}$B~\cite{Charity:2011}, $^{17}$Na~\cite{Brown:2017}, $^{31}$K~\cite{Kostyleva:2019}, and $^{13}$F~\cite {Charity:2021}. The measured 3\textit{p}-decay patterns in all cases include
2\textit{p} emission as part of a sequential \textit{p}--2\textit{p} decay mechanism. This may strongly influence the predictions for unobserved-yet isotopes.
More multi-proton decay modes are reported, i.e.\ 5\textit{p} emission from $^{9}$N~\cite{Charity:2023}), and even 6\textit{p} emission is foreseen from unobserved yet $^{20}$Si.
\\
iv)
Predictions for proton-unbound isotopes by using their neutron-rich mirrors and isospin symmetry indicate area of 5--6 \textit{amu} beyond the proton dripline~\cite{Grigorenko:2018}. A mirror symmetry emerged from the isobaric-spin formalism means that pairs of the same-mass nuclei with reversed numbers of protons and neutrons should have an identical set of states including their g.s.~ (i.e., with the same total angular momentum $J$ and parity $\pi$). For the most remote nuclear systems, no g.s.\ of isotopes (and thus no new isotope identification) are expected. Therefore a new borderline indicating the limits of existence of isotopes in the nuclear chart and the transition to chaotic-nucleon matter may be discussed~\cite{Kostyleva:2019}.

In the present work, we continue the ``excursion beyond the proton dripline'' of Ref.~\cite{Mukha:2018,Grigorenko:2018} by reporting the results of additional analysis of the data obtained with a $^{20}$Mg secondary beam~\cite{Mukha:2007}. The by-product data analysis is aimed at previously-unobserved  3\textit{p}-unbound  isotope $^{20}$Al produced in a charge-exchange reaction of the same $^{20}$Mg secondary beam. Then isospin symmetry of the $^{20}$Al--$^{20}$N mirror pair  is inspected.

The experiment was described in detail in Refs.~\cite{Mukha:2010,Mukha:2012}. The $^{20}$Mg beam was produced by the fragmentation of a primary 591 \textit{A}MeV $^{24}$Mg beam at the SIS-FRS facility at GSI (Germany). The main objective of the experiment was study of 2$p$-decays of $^{19}$Mg isotopes in flight. A brief summary of the experimental setup and detector performance are given below. The FRS was operated with ion-optical settings in a separator-spectrometer mode, where the first half of the FRS was set for separation and focusing of the radioactive beams on a secondary target in the middle of the FRS, and the second half of FRS was set for the detection of heavy-ion decay products. The secondary $^{20}$Mg beam with an energy of 450 \textit{A}MeV and an intensity of 400 ions $\text{s}^{-1}$ bombarded a 2 g/cm$^{2}$ $^9$Be secondary target located at the FRS middle focal plane, see details in Refs.~\cite{Mukha:2007,Mukha:2012}. The $^{19}$Mg nuclei were produced via a neutron-knockout reaction with the $^{20}$Mg projectiles. The decay products of unbound $^{19}$Mg nuclei were tracked by a double-sided silicon micro-strip detector (DSSD) array placed just downstream of the secondary target. Four large-area DSSDs~\cite{Stanoiu:2008} were employed to measure hit coordinates of the protons and the recoil heavy ions (HI), resulting from the in-flight decays of the studied 2$p$ precursors. The high-precision position measurement by DSSDs allowed for reconstruction of all fragment trajectories, enabling us to derive the decay vertex together with angular HI-$p$ and HI-$p-p$ correlations. For example, the spectroscopic information on $^{19}$Mg was obtained based on the analysis of the trajectories of $^{17}$Ne+\textit{p}+\textit{p} measured in coincidence~\cite{Mukha:2012}. Several states of $^{19}$Mg were observed by using 2\textit{p} angular correlations as a function of their root-mean-square angle relative to $^{17}$Ne,
\begin{equation}\label{eq:1}
\rho_\theta = \sqrt{\theta^2_{p_1-^{17}Ne}+\theta^2_{p_2-^{17}Ne}}.
\end{equation}
A number of by-product results were obtained in a similar way from the data recorded in the same experiment. In particular, a 3\textit{p}-unbound nuclear system of $^{20}$Al was populated in a charge-exchange reaction. Such a mechanism has smaller cross section than knockout reactions, thus the $^{20}$Al data have less statistics than the $^{19}$Mg data~\cite{Mukha:2012}. Nevertheless, they may serve as the first hints on nuclear structure of the previously-unobserved $^{20}$Al. The $^{20}$Al spectrum was derived from trajectories and angular correlations of all decay products $^{17}$Ne+$p_1$+$p_2$+$p_3$ which were measured in four-fold coincidence. All coefficients for detector calibrations were taken from the analysis reported in Refs.~\cite{Mukha:2010,Mukha:2012}, and the same data analysis procedure was employed for $^{20}$Al as that applied to 3\textit{p}-decays of $^{31}$K~\cite{Kostyleva:2019}.

The measured trajectories of $^{17}$Ne+3\textit{p} coincident events were used for deriving  relative angles between each proton and $^{17}$Ne. Then a kinematic variable $\rho_3 $ was introduced for 3\textit{p} decays in analogy with 2\textit{p} decays [see eq.~(\ref{eq:1})] and its expression is as follows: 
\begin{equation}
\rho_3 = \sqrt{\theta^2_{p_1-^{17}Ne}+\theta^2_{p_2-^{17}Ne}+\theta^2_{p_3-^{17}Ne}}
\label{eq:2}
\end{equation}
The decay protons share the total 3\textit{p}-decay energy $E_T$, thus the $\rho_3$ variable is very useful for illustration of the states in the 3\textit{p}-decay precursor, like is was shown in the case of $^{31}$K spectroscopy ~\cite{Kostyleva:2019}.

\begin{figure}[!t]
\begin{center}
\hspace{-4 mm}
\includegraphics[width=0.47\textwidth]{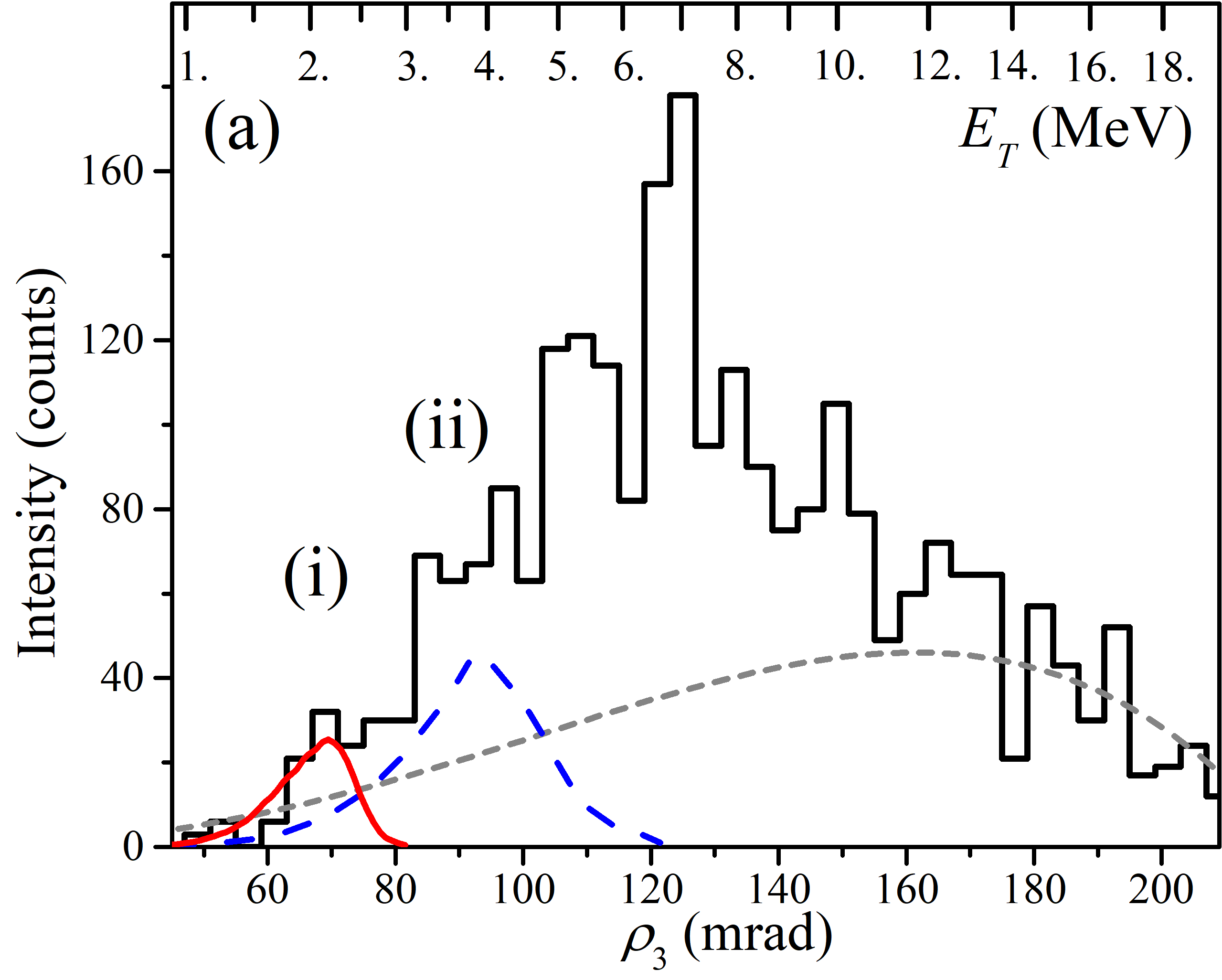}
\includegraphics[width=0.46\textwidth]{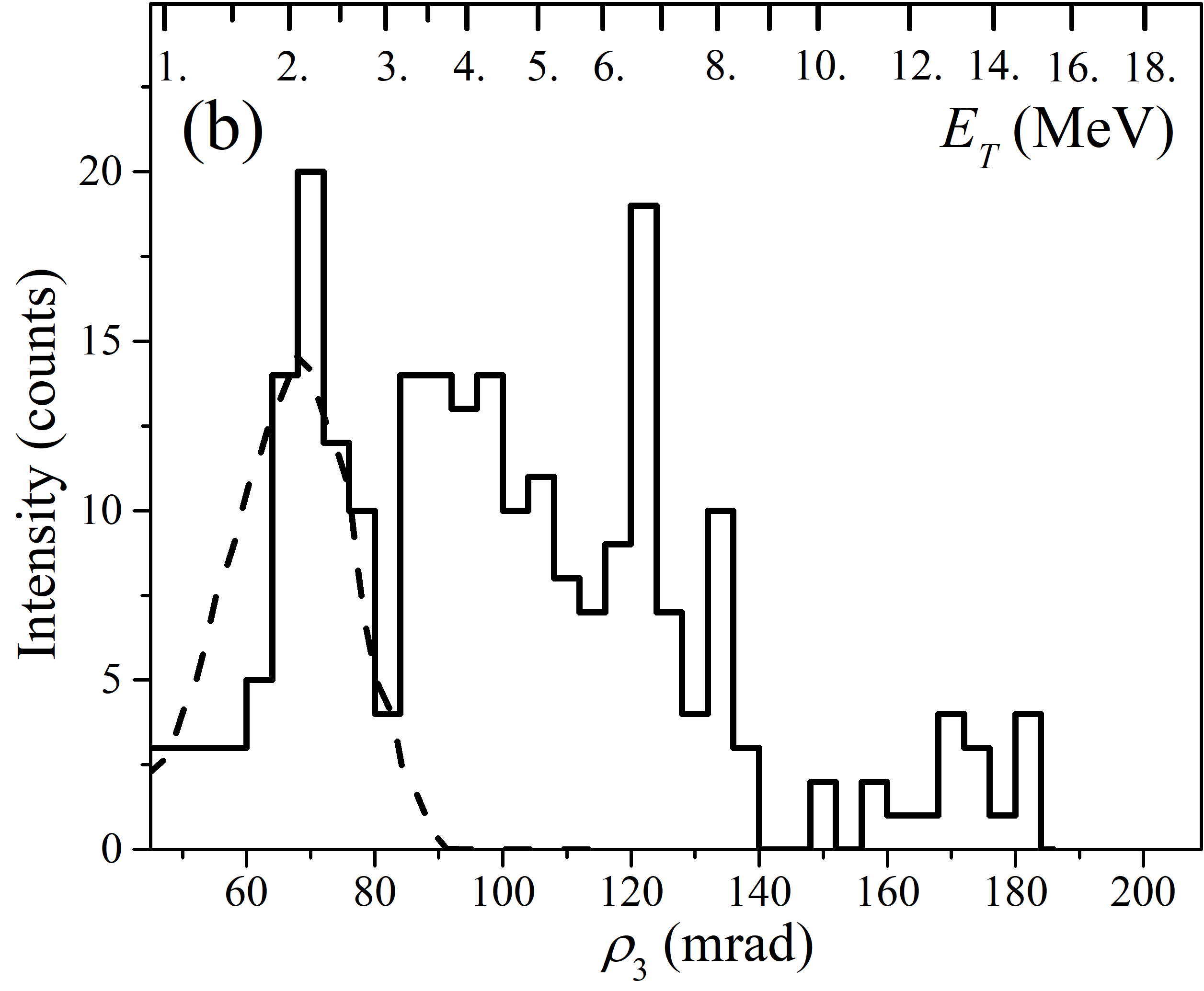}
\end{center}
\caption{(a) Three-proton angular correlations $\rho_3$ [see eq.~\ref{eq:2}] derived from the measured trajectories of all decay products, $^{17}$Ne+3$p$ (histogram), which reflect total 3$p$-decay energy $E_T$ of the $^{20}$Al states shown in the upper axis. {The labels (i) and (ii) indicate regions where populations of the two lowest states in $^{20}$Al are expected. The illustrative simulations of the $^{20}$Al 3$p$-decays with the assumed $E_T$  of 2.0 and 3.6 MeV are shown by the solid and dashed curves, respectively. The grey short-dashed curve shows a four-body phase volume simulations for a direct reaction with an exit channel $^{17}$Ne+3$p$ without any resonance in $^{20}$Al.} 
(b) The similar distribution to that shown in (a) but gated in addition by small relative angles 20$\leq\theta_{p-^{17}Ne}\leq$40 mrad, which is typical for the $^{19}$Mg g.s.\ decay (see the area under dash-dotted line in Fig.~\ref{fig:theta-p-Ne}). The dashed curve shows simulations of the $^{20}$Al 3$p$-decay with $E_T$=2.0 MeV.}
\label{fig:20Al-rho3}
\end{figure}

{
 The $\rho_3$ distribution derived from the measured $^{17}$Ne+3\textit{p} correlations following $^{20}$Al decays is shown in Fig.~\ref{fig:20Al-rho3}(a).
The corresponding total 3\textit{p}-decay energies $E_T$ can be estimated from the upper axis. One can see that the low-energy part cannot be described by a 4-body phase volume  simulated for a direct reaction with an exit channel $^{17}$Ne+3$p$ without any resonance in $^{20}$Al. The  4-body phase volume is
proportional to an $E_T^{7/2}$ factor multiplied to the detection efficiency of events $^{17}$Ne+3$p$.
It is normalized to the measured intensity at small and large $\rho_3$ values, $\le$60 and $\ge$160 mrad, respectively. 
Due to this normalization, the 4-body phase-volume component represents rather an upper-limit estimate of possible contribution of non-resonant branch in the measured correlations.
The data significantly exceed the estimated non-resonant contribution, and therefore low energy $^{20}$Al resonance contributions 
are required. 
The energy of the $^{20}$Al g.s.\ was predicted by the systematics proposed for the mass differences of mirror nuclei (the improved Kelson-Garvey mass relations~\cite{Tian:2013,Zhong:2022}).  The estimated $E_T$ values for  g.s.~ of $^{20}$Al are 3.4--3.6 MeV, which reduces the inspected 
angular correlations down to the range of 60--100 mrad, see Fig.~\ref{fig:20Al-rho3}(a).
There are few bumps in the $\rho_3$ distribution above 100 mrad, e.g., located at $E_T$ of $\sim$5 and  $\sim$7 MeV, which may correspond to higher excited states in $^{20}$Al and may be addressed elsewhere. In this work, we focus on search of the lowest states in $^{20}$Al.
} 

{
With the predicted $E_T$ values, $^{20}$Al g.s.\ should be open to a sequential 1\textit{p}--2\textit{p} decay mechanism via the intermediate g.s.\ of  $^{19}$Mg. The known decay pattern of the $^{19}$Mg g.s.\ shows the $\theta_{p-^{17}Ne}$ correlations ranging from 20 to 40 mrad \cite{Mukha:2012}.
Then the measured $\rho_3$-correlations   may be exclusively inspected by implementing a selecting  $\theta_{p-^{17}Ne}$ gate. We produced the exclusive $\rho_3$ distribution by applying the gate in the  $\theta_{p-^{17}Ne}$ range of 20--40 mrad, which is typical for the $^{19}$Mg g.s.\ decay. As a consequence of such a gated projection, the low-energy states in $^{ 20}$Al decaying via the $^{19}$Mg g.s.\ should be conserved in comparison to higher-energy states which are open to several decay channels in larger $\theta_{p-^{17}Ne}$ ranges and therefore should be suppressed. Fig.~\ref{fig:20Al-rho3}(b) shows the gated $\rho_3$ distribution where there are two prominent peaks at  $E_T$ of 2.0 and 3.6 MeV, which suggests two low energy states in $^{20}$Al.
For illustration purpose, two possible states in $^{20}$ assumed at  $E_T$ of 2.0 and 3.6 MeV are shown in Fig.~\ref{fig:20Al-rho3}(a). Their contributions  are obtained by the GEANT simulations of the setup response to the reactions of interest and the data analysis applied to $\rho_3$ angular correlations, see the corresponding description in Appendix~\ref{Q3p_value}. The simulated peak regions are labeled as (i) and (ii), and their possible contributions exceed the upper-limit estimates of the non-resonance branch.  Such a qualitative description calls for further quantitative analysis of the $^{17}$Ne+3$p$ correlation events. 
} 

\begin{figure}[!htbp]
\begin{center}
\includegraphics[width=0.48\textwidth]
{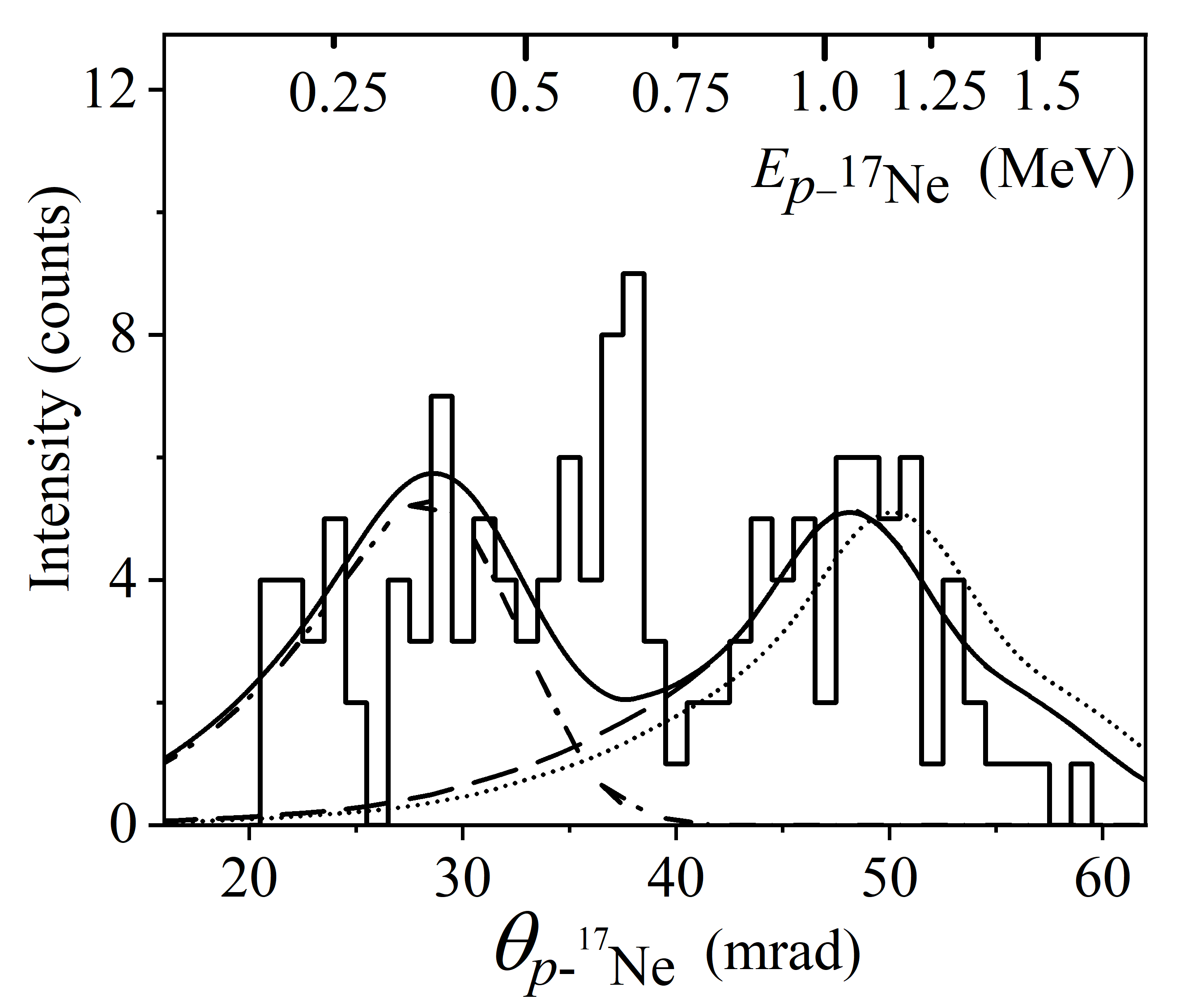}
\end{center}
\caption{Angular $\theta_{p-^{17}Ne}$ correlations (histogram) derived from the measured $^{17}$Ne+3$p$ coincidences by using the selection gate (i) in the 45$\le \rho_3 \le$82 mrad region corresponding to the bump at $E_T\sim2$ MeV in Fig.~\ref{fig:20Al-rho3}. The corresponding $1p$-decay energies $E_{p-^{17}Ne}$ are given by the upper axis. The simulated contribution from an initial 1\textit{p}-decay of $^{20}$Al into the $^{19}$Mg g.s.\ with the estimated 1\textit{p}-decay energy of 1.17($^{+0.10}_{ -0.08}$) MeV is shown by the dashed curve. The contribution of a subsequent 2\textit{p}-decay of $^{19}$Mg g.s.\ with the known decay energy of 0.76(5) MeV~\cite{Mukha:2012,Xu:2018} is shown by the dash-dotted curve. The solid curve is their sum corresponding to $Q_{3p}$=1.93($^{+0.11}_{ -0.09}$) MeV. For illustration of uncertainties of the derived energies, the contribution of initial 1\textit{p}-decay of $^{20}$Al into the $^{19}$Mg g.s.\ with the shifted by +0.1 MeV 1\textit{p}-decay energy is shown by the dotted curve.}
\label{fig:theta-p-Ne}
\end{figure}

{
In order to establish the low-lying states in $^{20}$Al and their decay schemes quantitatively, the events located around the $\rho_3$ regions (i) and (ii) in Fig.~\ref{fig:20Al-rho3}(a) were selected, and the respective angular $\theta_{p-^{17}Ne}$ correlations were examined. Fig.~\ref{fig:theta-p-Ne} displays the $\theta_{p-^{17}Ne}$ distribution by imposing the $\rho_3$ gate (i). The lowest-energy {bump} (i) around 2 MeV may correspond to the $^{20}$Al g.s.\ which decays by sequential emission of a proton into an intermediate state, $^{19}$Mg g.s.\ whose 2\textit{p}-decay energy of 0.76(5) MeV has been measured~\cite{Mukha:2012,Xu:2018}. Then the corresponding $\theta_{p-^{17}Ne}$ correlations in Fig.~\ref{fig:theta-p-Ne} should consist of two contributions. The first-emitted proton into an intermediate state in $^{19}$Mg is expected to cause a peak in the observed $\theta_{p-^{17}Ne}$ correlations. The second component should have the same shape as the known relatively broad $\theta_{p-^{17}Ne}$ distribution from the $^{19}$Mg g.s.\ 2\textit{p}-decay~\cite{Mukha:2012,Xu:2018}, which is centered around $E_{p-^{17}Ne}\simeq$ 0.38 MeV (because two identical protons share the total 2\textit{p}-decay energy of 0.76 MeV). We have evaluated the data distribution in Fig.~\ref{fig:theta-p-Ne} by a sum of two respective components: 1) the simulation of the response of the experimental setup to the 1\textit{p}-emission of $^{20}$Al (the simulation procedure is described in details in Refs.~\cite{Mukha:2010,Mukha:2012}); 2) the known detector response to the 2\textit{p}-decay of $^{19}$Mg g.s.\ (see Refs.~\cite{Mukha:2012,Xu:2018}). One may see that the small-angle region of the $\theta_{p-^{17}Ne}$ distribution is described by the 2\textit{p}-decay of $^{19}$Mg g.s.\ (the dash-dotted curve reflects the known value of the 2\textit{p}-decay energy of 0.76~MeV), while the large-angle correlations can be described by the 1\textit{p}-emission of $^{20}$Al into $^{19}$Mg g.s.\ (dashed curve) with the estimated 1\textit{p}-decay energy of 1.17($^{+0.10}_{ -0.08}$) MeV, which results in the total 3\textit{p}-decay energy $E_T$=1.93($^{+0.11}_{ -0.09}$) MeV. More details on deriving $E_T$ values may be found in Appendix~\ref{Q3p_value}.
} 

%


\begin{figure}[!htbp]
\begin{center}
\includegraphics[width=0.48\textwidth]{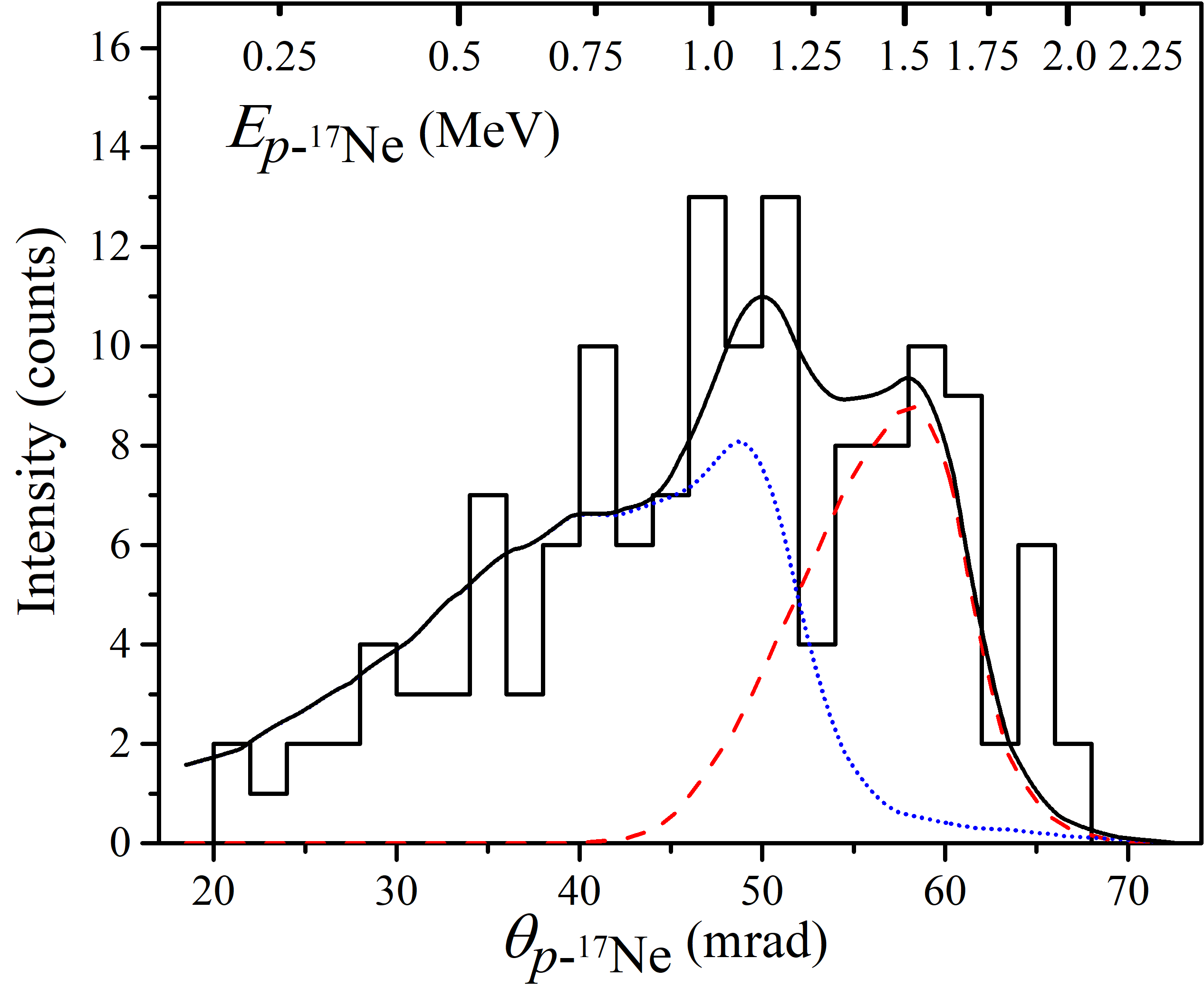}
\end{center}
\caption{Angular $\theta_{p-^{17}Ne}$ correlations (histogram) derived from the measured $^{17}$Ne+3$p$ coincidences by using the selection gate (ii) in the 82$\le \rho_3 \le$92 mrad range corresponding to the  {bump} at $E_{T}\simeq$3.5 MeV in Fig.~\ref{fig:20Al-rho3}. The corresponding $1p$-decay energies $E_{p-^{17}Ne}$ are given by the upper axis. The simulated contribution from the primary 1\textit{p}-decay of $^{20}$Al$^{*}$ into the first excited state in $^{19}$Mg$^{*}$ (at $Q_{2p}$=2.1(2) MeV) with the estimated 1\textit{p}-decay energy of 1.50(10) MeV is shown by the dashed curve. The contribution of secondary-emitted protons from the $^{19}$Mg$^{*}$ into the $^{17}$Ne g.s.\ measured in Ref.~\cite{Mukha:2012} is shown by the dotted curve. The solid curve shows their sum.}
\label{fig:theta-pNe_exc}
\end{figure}

{
The angular $\theta_{p-^{17}Ne}$ correlations obtained by selection using the $\rho_3$ gate (ii) in Fig.~\ref{fig:20Al-rho3} are shown in Fig.~\ref{fig:theta-pNe_exc}. Such a selection is aimed at an excited state in $^{20}$Al located around $E_T\simeq$3.6 MeV. One may see two bumps dominating the selected distribution at the angles of $\sim$50 and $\sim$58 mrad. These bumps are interpreted as due  to the sequentially-emitted protons from $^{20}$Al$^{*}$ via an intermediate state in $^{19}$Mg$^{*}$($Q_{2p}$=2.1(2) MeV) and its subsequent 2\textit{p} decay via $^{18}$Na states \cite{Mukha:2012}. The angular correlations from the 2\textit{p} decay of the intermediate 2.1-MeV state in $^{19}$Mg$^{*}$ have been measured and described in Ref.~\cite{Mukha:2012}. The respective simulation taken from Fig.~4(b) of Ref.~\cite{Mukha:2012} is shown in Fig.~\ref{fig:theta-pNe_exc} (the dotted curve normalized to the data).  The dotted curve is double-humped since it represents sequential $^{19}$Mg$^{*}\!\rightarrow^{18}$Na+$p\rightarrow^{17}$Ne+2\textit{p} decays as explained in Fig.~4(b) of Ref.~\cite{Mukha:2012}. The contribution of the first-emitted proton is shown by the dashed curve. The sum of these two contributions matches the data with probability of 0.993 when the energy of the first-emitted proton is 1.50($^{+0.12}_{ -0.10}$) MeV. The 1\textit{p}-decay energy and its uncertainties are derived similarly to those of the $^{20}$Al g.s., see Appendix~\ref{Q3p_value}. In sum with the  energy of the secondary- and tertiary- emitted protons $Q_{2p}$=2.1(2) MeV, the total 3\textit{p}-decay energy is $Q_{3p}$=3.60(22) MeV. 
} 

\begin{figure}[!htbp]
\begin{center}
\includegraphics[width=0.48\textwidth]{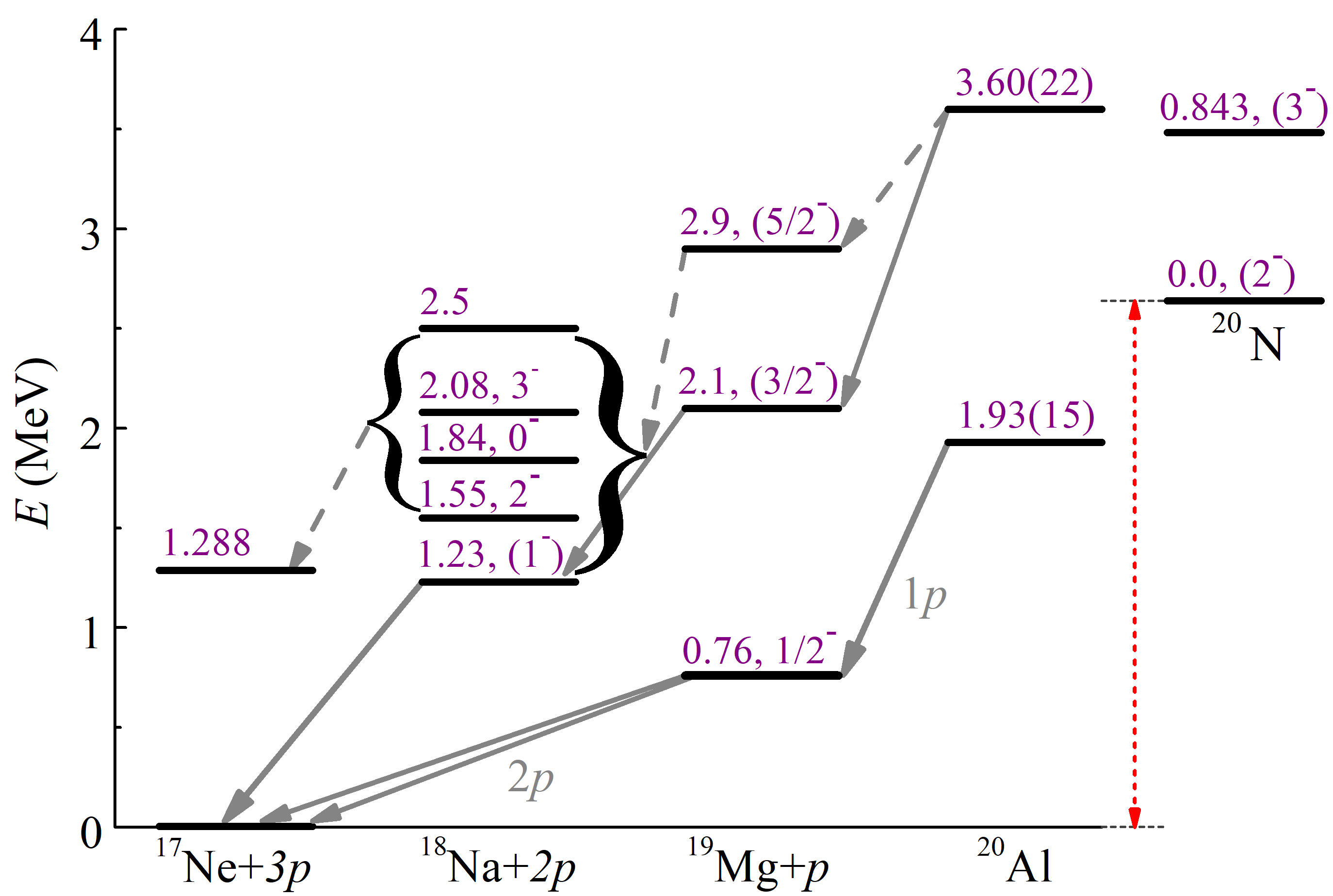}
\end{center}
\caption{Proposed decay scheme of the two lowest states in $^{20}$Al with tentatively-assigned 1\textit{p}-decay channels through the known $^{19}$Mg and $^{18}$Na ground states~\cite{Mukha:2012}, whose energies are given relative to the 3$p$, 2$p$ and 1$p$ thresholds, respectively. On the right-hand side, the levels of mirror nucleus $^{20}$N are shown. They are shifted by the mirror energy difference MED expected for the $s_{1/2}$ 1\textit{p} configuration in the $^{20}$N-$^{20}$Al pair, which corresponds to the closest prediction $S_{3p}$($^{20}$Al)=$-$2.64 MeV, see the vertical dashed arrow. The MED estimates are obtained by using the parameterization given in Ref.~\cite{Fortune:2018}.}
\label{fig:decay-scheme}
\end{figure}

The assigned energies of two low-lying states in $^{20}$Al and their decay scheme are displayed in Fig.~\ref{fig:decay-scheme}. The derived information may be improved in future experiments with higher statistics and better resolution. The mass excess of the $^{20}$Al g.s.\ has been derived by using the masses of $^{17}$Ne+3\textit{p} and its decay energy, which provides the value of +40.30(15) MeV.
The energy of the $^{20}$Al g.s.\ predicted by the improved Kelson-Garvey mass relations~\cite{Tian:2013,Zhong:2022} (the evaluated $S_{3p}$ value is $-$(3.4--3.6) MeV) is in a large disagreement with the data. Such a difference may be explained by the effect of Thomas-Ehrmann shift~\cite{Thomas:1952,Ehrman:1951} which is often observed in 1\textit{p}-unbound nuclei. Indeed, as the $^{20}$Al g.s.\ decays via the relatively long-lived $^{19}$Mg g.s., one may use the empirical $S_p$ systematics derived from the known 1\textit{p}-emitting states in light nuclei. It is based on a parameterization of the mirror energy difference MED~\cite{Fortune:2018}. The definition of MED is MED=$S_n$(neutron-rich nucleus)$-S_p$(its proton-rich mirror), and the parametrization is MED=(\textit{Z/A}$^{1/3}$)MED$^\prime$, where the MED$^\prime$ value does not depend on the proton number \textit{Z} and mass number \textit{A}~\cite{Fortune:2018}. We evaluated the $S_{3p}$ value for the $^{20}$Al g.s.\ by using the known $S_{n}$ value of its mirror partner $^{20}$N$_{g.s.}$(2$^{-}$) and the corresponding MED value taken from the parameterization~\cite{Fortune:2018}. If the $^{20}$Al g.s.\ is an \textit{s}-wave state, the predicted 3\textit{p}-separation energy is $S_{3p}$($^{20}$Al)=-2.64 MeV which is higher than the data, see Fig.~\ref{fig:decay-scheme}. If the $^{20}$Al g.s.\ is a $d_{5/2}$ state, the parameterization results in $S_{3p}$($^{20}$Al)=-3.42 MeV. This predicted value is again considerably higher than the data.~Thus we conclude that the Thomas-Ehrmann shift (which is well studied in 1\textit{p}-emitters) may partly explain the observed lowering of 3\textit{p}-unbound g.s.\ of $^{20}$Al in comparison with its bound isospin mirror $^{20}$N, though the observed lowering effect is significantly larger.

\begin{figure}[!htbp]
\begin{center}
\includegraphics[width=0.48\textwidth]{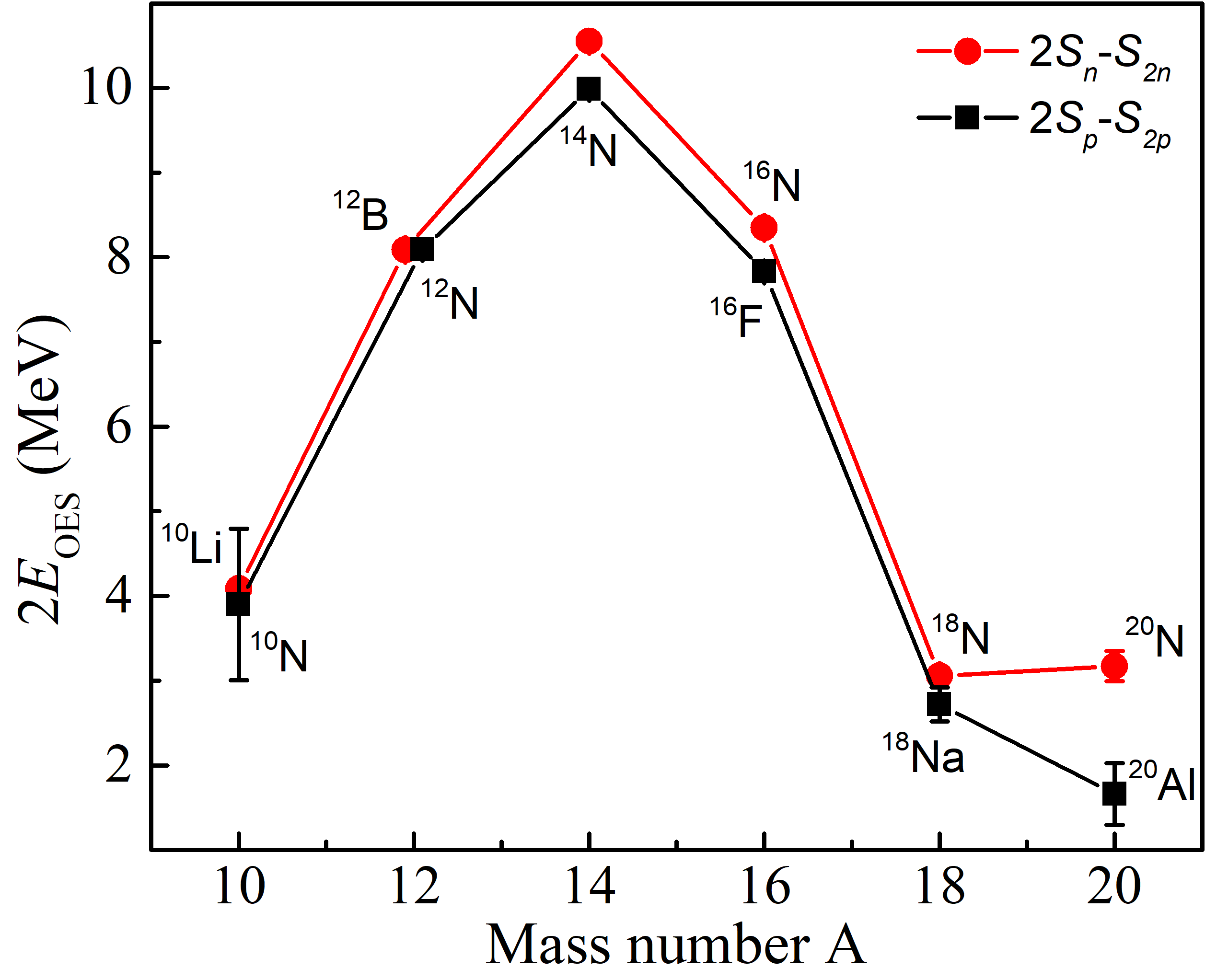}
\end{center}
\caption{The OES in nuclear masses for nitrogen isotopes and their mirror nuclei. The red solid circles represent the OES values calculated by the expression $2E_{OES}=2S_n-S_{2n}$, while the black solid squares denote those calculated by the expression $2E_{OES}=2S_p-S_{2p}$. The data points for $^{12}$N and $^{12}$B are partially shifted in the x-axis to avoid overlapping.}
\label{fig:Eoes_N}
\end{figure}

{
 Based on the obtained mass of $^{20}$Al, we can study the systematics of odd-even staggering (OES) of nuclear masses, which were proven to be a helpful indicator on mirror energy differences in our previous studies~\cite{Mukha:2015,Mukha:2018}. The OES is defined as $2E_{OES}$=$2S_N$--$S_{2N}$. Here, $S_N$ and $S_{2N}$ are one-nucleon (either proton or neutron) and two-nucleon separation energies, respectively. The systematics of OES energy for nitrogen isotopes and their mirror nuclei is presented in Fig.~\ref{fig:Eoes_N}. One can see that the $E_{OES}$ of proton-rich nuclei is always smaller than that of neutron rich partner. The difference is small except for the mirror pair $^{20}$N-$^{20}$Al. The corresponding value for $E_{OES}$ difference for $^{20}$N-$^{20}$Al is 0.755 MeV, which clearly indicates a Thomas-Ehrmann shift in this bound-unbound mirror pair.
}

\begin{figure}[!htbp]
\begin{center}
\includegraphics[width=0.48\textwidth]{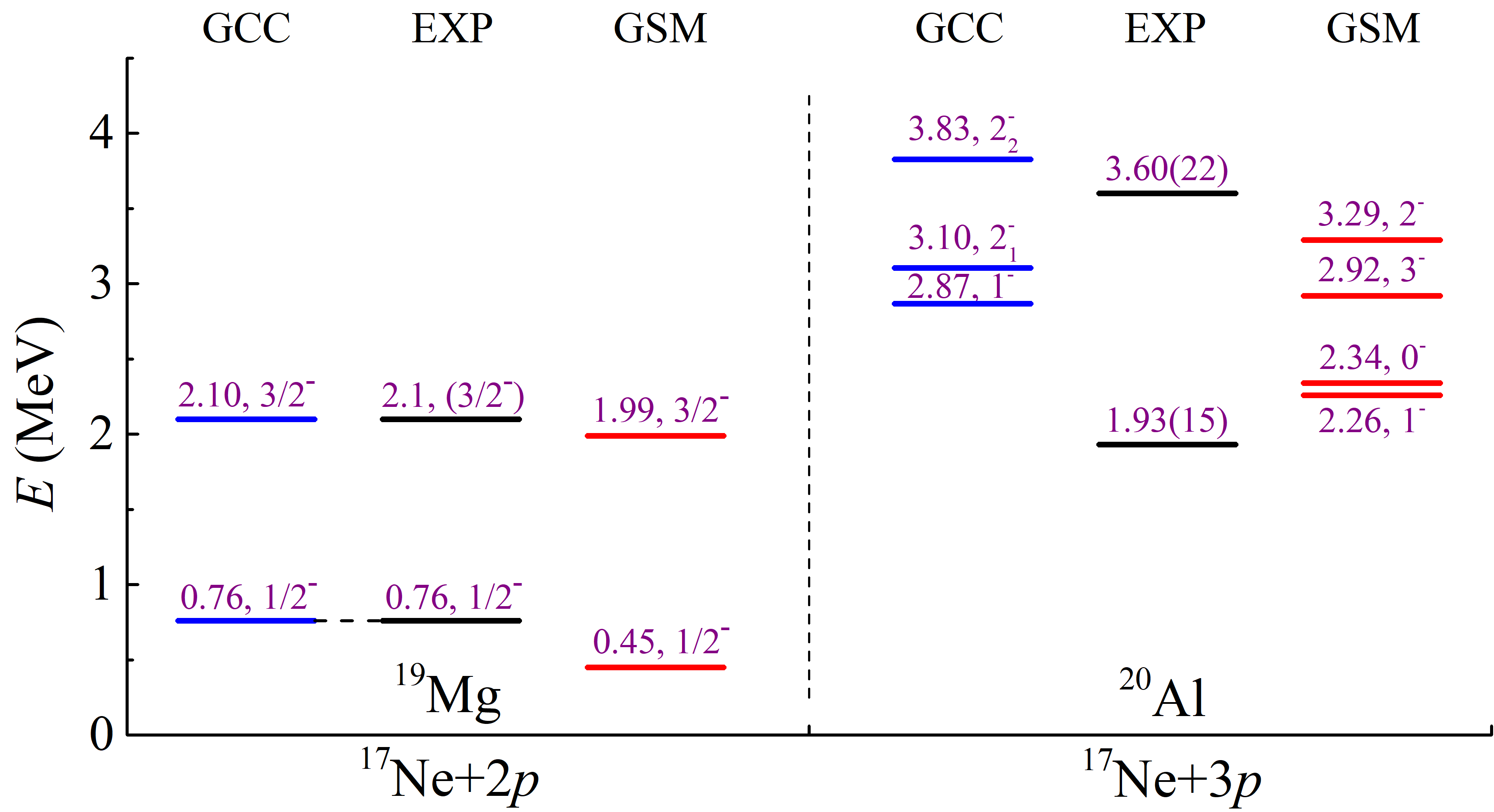}
\end{center}
\caption{The energies of lowest states in $^{20}$Al and those of $^{19}$Mg derived from the experiment compared to two model predictions. The energy values are provided with respect to the 3\textit{p} and 2\textit{p} thresholds, respectively.}
\label{fig:exp_th_comp}
\end{figure}

{
We calculated low lying states of $^{20}$Al and its 1\textit{p}-decay daughter nucleus $^{19}$Mg by employing the Gamow Shell Model (GSM)~\cite{Michel:2002,Michel:2009,Li:2021} and Gamow-Coupled-Channel (GCC) model~\cite{Wang:2017,Wang:2021,Wang:2022}. A brief description the models and corresponding calculations details can be found in Appendix~\ref{GSM_model} and Appendix~\ref{GCC_model}. Figure~\ref{fig:exp_th_comp} compares the model predictions with the data. The GSM provides a reasonable description of both $^{19}$Mg ${1/2}^-$ g.s.\ and $({3/2}^-)$ excited state, in particular, the energy of $({3/2}^-)$ state is well-reproduced within the experimental uncertainties. In the GCC model calculations, the energies of ${1/2}^-$ and $({3/2}^-)$ states in $^{19}$Mg were employed to adjust the free parameters. Its predictions for $^{20}$Al indicate that the g.s.\ has a spin-parity of $1^-$ and is located 2.87 MeV above the 3\textit{p} threshold. The GSM also predicts $J^\pi$=$1^-$ for the $^{20}$Al g.s., which is positioned at 2.26 MeV above the 3\textit{p} threshold. These results demonstrate that the $^{20}$Al g.s.\ has a dominant $s_{1/2}$ 1\textit{p}-configuration, which differs from the mirror $^{20}$N g.s.\ with the $J^\pi$=($2^-$). In addition to the energy shift of the $^{20}$Al g.s.\ relative to the $^{20}$N g.s., the spin-parity difference provides further evidence of mirror symmetry broken in $^{20}$Al-$^{20}$N pair. Regarding the excited states of $^{20}$Al, several low-energy levels are predicted by GSM. The  level predicted to be closest to the observed 3.6 MeV state is situated at 3.29 MeV above the 3\textit{p} threshold and has the $J^\pi$=$2^-$. The GCC model predicts two excited states with a spin-parity of $2^-$, and the one close to the 3.6 MeV state in terms of energy is located at 3.83 MeV above the 3\textit{p} threshold. Therefore, $J^\pi$=($2^-$) may be tentatively assigned to the observed 3.6 MeV state.
}

The discussed difference in the observed and predicted energies of the $^{20}$Al g.s.\ is not unique. Similar lowering of g.s.\ energy has been observed in another 3$p$-emitter $^{31}$K with a likewise ($d_{5/2}$)$^3$ configuration, which may indicate additional binding stemming from a nuclear structure effect that requires further investigation, such as employing the three-body forces~\cite{Holt:2013}. The data on other 3$p$-emitters, e.g.\ the unidentified-yet g.s.\ of $^{13}$F or $^{17}$Na with a presumed ($sd$)$^3$ configuration are also in request.

We have estimated the half-life values for the observed $^{20}$Al states by measuring distributions of their decay vertexes in the same way as in the previous study of 3\textit{p} decay of $^{31}$K~\cite{Kostyleva:2019}. All vertices are located within the reaction target, and therefore we found no indication on long-lived states in $^{20}$Al. The width of the $^{20}$Al g.s.\ derived by the description in Fig.~\ref{fig:theta-p-Ne} provides only the upper-limit value $\Gamma_{g.s.}<$ 400~keV, which is mainly due to the experimental resolution. For comparison, the upper-limit Wigner estimate for a single-particle 1$d_{3/2}$-shell width of the $^{20}$Al g.s.\ is about 30~keV only. The excited 3.6 MeV state is not detected to decay via the most energetically preferred branch into the $^{20}$Al g.s., which may be explained by a high orbital momentum $\ell_p$ required for the transition. Indeed, if we assume its $J^\pi$=4$^{-}$ (which may correspond to the 0.944-MeV state in mirror $^{20}$N) then $\ell_p$=4 and $\ell_p$=2 are required by the momentum conservation for the decays into the 1/2$^{-}$ g.s.\ and excited (3/2$^{-}$,5/2$^{-}$) states in $^{19}$Mg. In this way, a partial decay width of the 1\textit{p}-decay branch from $^{20}$Al$^{*}$(3.6 MeV) into $^{19}$Mg g.s.\ may be reduced by a factor of 1000 in comparison with the dominant branches.

The spectrum of $^{20}$N, which is the mirror nucleus of $^{20}$Al, exhibits several low-energy levels with excitation energies below the 1\textit{n}-threshold of 2.16 MeV. Therefore the broad peak (ii) in Fig.~\ref{fig:20Al-rho3} assigned as the first-excited state in $^{20}$Al may consist of two or more unresolved peaks due to excitation of other low-energy states.

The observation of the 3\textit{p}-unbound $^{20}$Al g.s.\ leads to the prediction that its neighboring isotope $^{21}$Si is a 4\textit{p} emitter. Taking into account that the predicted $S_{2p}$($^{21}$Si)=-3.7 MeV~\cite{Fortune:2017}, one may derive its $S_{p}$=-2.53 MeV and therefore to expect a triple sequential 1\textit{p}--1\textit{p}--2\textit{p} emission from the unobserved-yet g.s.\ of $^{21}$Si.

In conclusion, the first spectroscopy of the previously-unknown nucleus $^{20}$Al which decays by 3\textit{p} emission, has revealed low-energy states with energies significantly lower than the anticipated values derived from their isobaric mirror partners. The mass excess of the $^{20}$Al g.s.\ derived from the measured $S_{3p}$ value and all fragment masses is +40.30(15) MeV, which is a challenging test of predictions of nuclear mass models. The observed effect of increased Thomas-Ehrman shift of the 3\textit{p}-unbound g.s.\ of $^{20}$Al can be theoretically explained by both GSM and GCC model calculations where an $s$-wave component of valence protons is dominant, which results also in the predicted $J^\pi$=1$^-$. This indicates a broken isospin symmetry with the (2$^-$) g.s.\ of mirror $^{20}$N.
The observed effect of lowering of the $^{20}$Al states is similar to that detected in another 3$p$-emitter $^{31}$K, which indicates a possible phenomenon of nuclear-structure preservation far beyond the proton drip line, thereby calling for further systematic investigations. If the effect of nuclear-structure preservation is confirmed, then the region of existence of proton(s) resonances in nuclear chart is more broad in comparison with the previous estimates based on the isospin symmetry, i.e.\ the number of unknown-yet isotopes is larger, and the transition region to chaotic nuclear systems is located more far from the proton drip line in comparison with the previous predictions~\cite{Grigorenko:2018}.

This work was partially supported by the Helmholtz International Center for FAIR (HIC for FAIR); the Chinese Academy of Sciences President’s International Fellowship Initiative (Grant No. 2024PVA0005); the National Key Research and Development Program (MOST 2022YFA1602303 and MOST 2023YFA1606404), the National Natural Science Foundation of China (No. \,12347106, No. \,12147101, No. \,12205340); the Gansu Natural Science Foundation under Grant No. 22JR5RA123; the European Union’s Horizon Europe Research and Innovation program under Grant Agreement No. 101057511 (EURO-LABS); the Helmholtz Association (grant IK-RU-002); the Russian Science Foundation (Grant No. 22-12-00054); the Polish National Science Center (Contract No. 2019/33/B/ST2/02908); the Helmholtz-CAS Joint Research Group (grant HCJRG-108); the Ministry of Education \& Science, Spain (Contract No.\ FPA2016-77689-C2-1-R); the Ministry of Economy, Spain (grant FPA2015-69640-C2-2-P); the Hessian Ministry for Science and Art (HMWK) through the LOEWE funding scheme; the Justus-Liebig-Universit\"at Giessen (JLU) and the GSI under the JLU-GSI strategic Helmholtz partnership agreement; and DGAPA-PAPIIT IG101423. This work was carried out in the framework of the Super-FRS Experiment Collaboration.
\begin{appendix}
\section{Deriving the $^{20}$Al decay energies}\label{Q3p_value}
{
In order to determine the 3\textit{p}-decay energies $Q_{3p}$ of $^{20}$Al 
corresponding to the $\theta_{p-^{17}Ne}$ angular correlations in Fig.~\ref{fig:theta-p-Ne}, 
we performed Monte Carlo simulations
of the detector response to the sequential 1\textit{p}-2\textit{p} decays of $^{20}$Al  states via intermediate $^{19}$Mg by
using the GEANT software \cite{Agostinelli:2003}, which was described in detail in the previous evaluations of data \cite{Mukha:2010,Mukha:2012}.  For a large-angle peak in Fig.~\ref{fig:theta-p-Ne}, a number of simulations of angular correlations $^{17}$Ne-\textit{p}  were performed. Each simulation contained two components, one component with a varied  energy of 1\textit{p}-decay $^{20}$Al$\rightarrow^{19}$Mg, and another component with a fixed 2\textit{p}-decay energy of $^{19}$Mg 
($Q_{2p}$=0.76 MeV). The intrinsic widths of  $^{20}$Al states
were assumed to be very small, i.e., 1 keV. Then every simulated
spectrum was compared with the data by using the standard Kolmogorov test, which computes the probability that the simulated spectrum
matches the respective experimental pattern \cite{Eadie:1971}. According
to the Kolmogorov test, two compared histograms are statistical
variations of the same distribution if the Kolmogorov-test
probability value is larger than 0.5. The $Q_{1p}$ values were derived
from the distributions of the calculated probabilities with
the corresponding uncertainty. Consequently, the $Q_{1p}$ value
was determined to be 1.17$^{+0.10}_{ -0.08}$ MeV, which
corresponds to descriptions of the correlations in Fig.~\ref{fig:theta-p-Ne}
with the highest probability of 0.82.
}

{
The $\theta_{p-^{17}Ne}$  angular distribution in Fig.\ref{fig:theta-p-Ne} has two irregular bins around 38 mrad (the corresponding $E_p\simeq0.65$ MeV). Their sum  intensity is estimated to be comparable with the $\sim\!3\sigma$ statistical deviation of the contribution of the previously assumed 3\textit{p}-decay branch at the 38 mrad angle. Given the fact that the events of this ``peak'' are within the $\rho_3$ gate (i), this peak likely indicates another minor 3\textit{p}-decay channel with the total decay energy $E_T$=1.93 MeV. However, there are no excited states known (and/or predicted) in the intermediate $^{19}$Mg and $^{18}$Na nuclei which could match these $Q_{3p}$ and $E_p$ values. These events are consistent with population of the 3/2$^-$, E$^*$=1.288 MeV state in the final/ residual nucleus $^{17}$Ne.  The 1.288 MeV state of $^{17}$Ne de-excites by emitting a $\gamma$ ray, which is undetected in our experiment. Then a sum of the detected and undetected decay energies should match the observed excited state in $^{20}$Al at 3.60(22) MeV within statistical uncertainties. In such a plausible scenario, a minor branch of 3\textit{p} decay of the $^{20}$Al$^{*}$(3.6 MeV) state proceeds via three sequential proton emissions $^{20}$Al$^{*}\rightarrow \! p+^{19}$Mg$^{*}$($E_{2p}$=2.9 MeV)$\rightarrow \! 2p+ ^{18}$Na$^{*} \rightarrow \! 3p+ ^{17}$Ne$^{*}$($E^ *$=1.288 MeV), where intermediate state(s) in $^{18}$Na are unresolved. Then the energy of first-emitted proton from $^{20}$Al$^{*}$ to $^{19}$Mg$^{*}$($E_{2p}$=2.9 MeV) corresponds to the observed 0.65 MeV peak in Fig.~\ref{fig:theta-p-Ne} if $Q_{3p}$=3.55 MeV. However, contributions of other emitted protons are unresolved.
}


\section{Gamow shell model}\label{GSM_model}
{
The Gamow Shell Model utilizes the one-body Berggren basis~\cite{Berggren:1968}, which includes bound, resonant, and scattering states. In GSM, many-body correlations are incorporated through configuration mixing, while continuum coupling is inherently accounted for at the basis level~\cite{Michel:2002,Betan:2002,Michel:2009,Michel_GSM_book,Li:2021}. This allows GSM to effectively treat both continuum coupling and inter-nucleon correlations, making it a reliable predictive tool for describing weakly bound and unbound states, as demonstrated in Refs.~\cite{Michel:2022,Michel:2002,Michel:2021}.
GSM typically operates within a core-plus-valence-particle framework. In the present calculations, the interaction between the core and valence nucleons is modeled using a one-body Woods-Saxon potential, while the nucleon-nucleon interaction among the valence nucleons is described by effective field theory forces \cite{Machleidt:2011,Epelbaum:2009}, which have been applied in previous studies, such as the four-proton decays in $^{18}$Mg~\cite{Michel:2021,Jin:2021,Li_prc:2021}.
For the calculations of $^{20}$Al, we adopt the doubly magic nucleus $^{14}$O as the inner core, and use the Hamiltonian from Ref.~\cite{Michel:2021}, which was originally constructed for calculations involving $^{16}$Ne and $^{18}$Mg. Additionally, the leading-order $T=0$ channels (V$_s$ or V$_t$) are constrained to reproduce the ground state energy of $^{17}$Ne.
In the actual calculations, we first perform computations in the Berggren basis, where at most two nucleons (including valence protons and neutrons) are allowed to occupy scattering states, generating a natural orbital basis. Then, physical quantities are calculated within this natural orbital basis, where up to three protons or neutrons can occupy scattering states.
}

\section{Gamow Coupled-Channel model}\label{GCC_model}
{
In the Gamow-Coupled-Channel approach~\cite{Wang:2017,Wang:2021,Wang:2022},  $^{20}$Al is considered as a system comprising a deformed core, $^{18}$Mg, alongside a valence proton and neutron. Although this framework does not fully capture the unbound nature of this 3\textit{p} emitter, we focus on the structural configuration of the valence nucleons. The relative motion between the valence nucleons and the core is described through Jacobi coordinates~\cite{Wang:2017} with the Berggren ensemble~\cite{Berggren:1968}. To eliminate the Pauli-forbidden states, we apply a supersymmetric transformation~\cite{Sparenberg:1997,Wang:2018}.
}
{
For the interaction between the valence proton and neutron, we utilize the original Minnesota potential~\cite{Thompson:1977}. The core-valence nuclear potential adopts a Woods-Saxon (WS) form with the “universal” parameter set~\cite{Cwiok:1987} and a quadrupole deformation $\beta_2=-0.2$. To match experimental spectrum of $^{19}$Mg, we adjusted the depth of the WS central potential and the strength of its spin-orbit component to -43.9 MeV and 14.3 MeV, respectively. Additionally, we assume a dilatation-analytic form for the Coulomb potential between the core and the valence proton~\cite{Saito:1977}.
}
The calculations have been performed within a model space constrained by max$(l_{x}, l_{y})\le7$, and the maximal hyperspherical quantum number $K_{\mathrm{max}}=16$. To study resonances, the Berggren basis is applied to channels with $K\leq3$, while for higher $K$ channels, a harmonic oscillator basis with oscillator length b=1.75 fm and $N_{max}$=40 is employed. The complex-momentum contour for the Berggren basis spans $k = 0 \rightarrow 0.3-0.02i \rightarrow 0.5 \rightarrow 1.2 \rightarrow$ 6 fm$^{-1}$, being divided into 40 scattering states for each segment.

\end{appendix}

\bibliographystyle{apsrev4-2}
\bibliography{all}


\end{document}